# Glass-like thermal conductivity in SrTiO$_3$ thermoelectrics induced by A-site vacancies

S. R. Popuri,[a] A. J. M. Scott,[a] R. A. Downie,[a] M. A. Hall,[a] E. Suard,[b] R. Decourt,[c] M. Pollet,[c] and J.-W. G. Bos[a]*

[a] *Institute of Chemical Sciences and Centre for Advanced Energy Storage and Recovery, School of engineering and physical sciences, Heriot-Watt University, Edinburgh, EH14 4AS, United Kingdom. *j.w.g.bos@hw.ac.uk*

[b] *Institut Laue-Langevin, Grenoble, F-38000, France*

[c] *CNRS, Université de Bordeaux, ICMCB, 87 avenue du Dr. A. Schweitzer, Pessac F-33608, France*

**The introduction of A-site vacancies in SrTiO$_3$ results in a glass-like thermal conductivity while Nb substituted samples maintains good electrical conductivity. This unexpected result brings SrTiO$_3$ one step closer to being a high-performing phonon-glass electron-crystal thermoelectric material.**

**Introduction:**

Thermoelectric waste heat recovery is widely expected to be an important component of a sustainable energy future.[1] However, affordable and relatively high-performance materials are lacking. Metal oxides are good candidates because of their abundance, low toxicity, and stability at high temperatures.[2] In addition, metal oxides are used in multilayer capacitors and the infrastructure for large scale device production exists.[3] SrTiO$_3$ is among the most promising n-type materials because it has the unusual combination (for an oxide) of good electrical conduction ($\sigma$) and a high Seebeck coefficient (S), yielding power factors $S^2\sigma \leq 3.5$ mW m$^{-1}$ K$^{-2}$ in single crystals and epitaxial films.[4, 5] Unfortunately, the overall performance is compromised by a large thermal conductivity $\kappa$ = 12 - 6 W m$^{-1}$ K$^{-1}$ for undoped SrTiO$_3$ (Fig. 1),[6] which reduces the thermoelectric figure of merit, ZT = (S$^2\sigma$/$\kappa$)T, where T is the absolute temperature. Outstanding thermoelectric materials generally have phonon-glass and electron-crystal (PGEC) properties,[7] meaning that the electronic transport is characteristic of a crystalline solid, while the $\kappa$ is low and resembles that of a glass. The main focus in the optimisation of $SrTiO_3$ has therefore been on the reduction of $\kappa$, which consists of a large lattice contribution ($\kappa_{lat}$) and a small electronic component ($\kappa_{el} = \sigma LT$; L is the Lorenz number). One approach to reduce $\kappa_{lat}$ is to introduce point defects within the perovskite structure. This is for example used in $Sr_{1-x}La_xTiO_{3-\delta}$ ($0 \leq x \leq 0.15$) and $SrTi_{1-y}Nb_yO_{3-\delta}$ ($0 \leq y \leq 0.2$) and results in $\kappa_{lat} \approx 3$ W m$^{-1}$ K$^{-1}$ at 1073 K and a maximum ZT = 0.35 at 1073 K.[5] However, to achieve ZT = 1, a much more substantial reduction to $\kappa$ = 1-2 W m$^{-1}$ K$^{-1}$ is needed. One drawback of $La^{3+}$ and $Nb^{4+}$ substitution is that the conflicting requirements of charge carrier doping limit x, y $\leq 0.2$. However, even for $Sr_{1-x}Eu_xTi_{0.8}Nb_{0.2}O_{3-\delta}$ with $Sr^{2+}/Eu^{2+}$ mixtures ($0 \leq x \leq 1$), $\kappa$ was not reduced below ~3 W m$^{-1}$ K$^{-1}$ for x = 0.5 at 1000 K.[8] Recently, perovskites with A-site vacancies have started to generate attention. Promising $S^2\sigma$ values and low $\kappa$ were reported in A-site and oxgyen deficient $Sr_{1-x}Pr_{0.67x}TiO_{3-d}$ and $Sr_{1-x}Ti_{0.8}Nb_{0.2}O_{3-\delta}$,[9] while reductions in $\kappa$ were also observed in $Ca_{1-x}Nd_{0.67x}MnO_{3-\delta}$ perovskites.[10] Here, we report a systematic investigation of the A-site deficient $Sr_{1-x}La_{0.67x}\square_{0.33x}Ti_{1-y}Nb_yO_{3-\delta}$ perovskites. In order to separate the impact of A-site vacancies from oxygen defects we first prepared the electrically insulating oxygen stoichiometric $Sr_{1-x}La_{0.67x}\square_{0.33x}TiO_3$ series and measured its thermal conductivity. Compositions with x = 0, 0.4 and 0.8 were synthesised using standard solid state reactions in air and have lattice parameters in good agreement with literature values (see supplementary Table S1).[11] The diffraction patterns do not show any evidence for the presence of impurities (see supplementary Fig. S1), while the sample densities are 92-97% of the theoretical values (Table S1).

**Synthesis and Experimental details:**

[‡]Polycrystalline $Sr_{1-x}La_{0.67x}\square_{0.33x}TiO_3$ (x = 0, 0.4 and 0.8) samples were prepared on a one gram scale by heating cold pressed pellets containing ground mixtures of $SrCO_3$, $La_2O_3$ and $TiO_2$ at

1200°C for two times 12 hours, and at 1400°C for 4 hours in air, with intermediate regrinding between steps. 1% excess $SrCO_3$ was used for the x = 0.4 and x = 0.8 samples. The Nb doped samples were prepared using a similar procedure but all heating steps were done under 5% $H_2$ in $N_2$. $Nb_2O_5$ was used as starting material. The final heating step for the $Sr_{1-x}La_{0.67x}\square_{0.33x}Ti_{0.8}Nb_{0.2}O_{3-\delta}$ and $Sr_{0.80}La_{0.13}\square_{0.07}Ti_{1-y}Nb_yO_{3-\delta}$ series was 1450 °C. All samples were cooled naturally. Laboratory X-ray powder diffraction data were collected on a Bruker D8 Advance diffractometer with Cu $K_{\alpha 1}$ radiation. Neutron powder diffraction measurements were done on a three gram sample of $Sr_{0.80}La_{0.133}\square_{0.067}Ti_{0.95}Nb_{0.05}O_{3-\delta}$ using the super-D2B instrument at the Intitut Laue Langevin, Grenoble, France. The wavelength $\lambda = 1.594$ Å and data were binned in 0.05° steps between 5-160°. Rietveld fits were performed using the GSAS/ EXPGUI programmes.[18] Seebeck and electrical resistivity data were collected using a Linseis LSR-3 instrument. Thermal diffusivity ($\alpha$) and heat capacity ($C_p$) were measured using a Netzch LFA 457 and Perkin Elmer DSC 8500, respectively. The thermal conductivity was calculated using: $\kappa = \alpha(T)C_p(T)\rho$, where $\rho$ is the density. The lattice thermal conductivity was estimated using the Wiedemann-Franz law: $\kappa_{lat} = \kappa - L\sigma T$, where L is the Lorenz number ($2.44 \times 10^{-8}$ $V^2$ $K^{-2}$).

**Results and Discussions:**

The temperature dependence of $\kappa$ is given in Fig. 1, while the measured thermal diffusivity ($\alpha$) and heat capacity ($C_p$) data are given in Fig. S2. The $\kappa(T)$ is rapidly suppressed with increasing vacancy concentration, and becomes almost temperature independent for x = 0.8 (27% A-site vacancies). An impressive overall reduction of 80% at 323 K and 60% at 873 K is observed. It is well established that a number of phonon-scattering mechanisms contribute to the reduction of $\kappa_{lat}$. These include point-defect, phonon-phonon and interface scattering.[12]

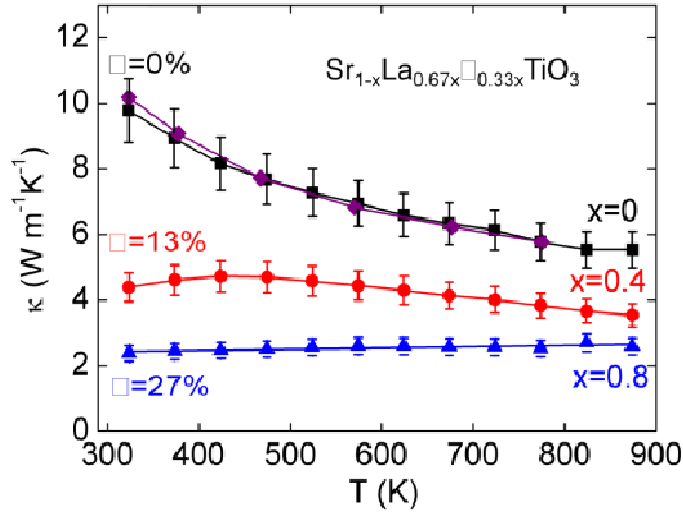

**Fig. 1.** Temperature dependence of the thermal conductivity (κ) for the $Sr_{1-x}La_{0.67x}\square_{0.33x}TiO_3$ series. (Literature data for $SrTiO_3$ (diamonds) is included for comparison).[6]

In crystalline solids, boundaries and point defects limit the thermal transport at low temperatures, while phonon-phonon Umklapp scattering dominates above the Debye temperature ($\theta_D$). This leads to the characteristic $1/T$ dependence at high-temperatures that is evident for $x = 0$ (Fig. 1), and for the A- and B-site doped $SrTiO_3$ compositions in the literature.[5, 8] The observation of an almost temperature independent $\kappa(T)$ for $x = 0.8$ is therefore of great interest. There is some precedent for this behaviour in perovskites, including in relaxor ferroelectrics and in segregated mixtures of $Nd_{0.67}TiO_3$ and $Nd_{0.5}Li_{0.5}TiO_3$.[13] A linear fit yields a positive slope ($3.9(8) \times 10^{-4}$ W m $K^{-2}$), which is characteristic of a glass with a constant phonon mean free path ($l_{ph}$).[12] An estimate of $l_{ph}$ (ignoring wavelength and frequency dependence) can be obtained from $\kappa_{lat} = 1/3 C_p v l_{ph}$ where $v$ is the velocity of sound which can be obtained from $\theta_D = (\hbar v/k_B)(6\pi^2 N)^{1/3}$, where $N$ is the number of atoms per unit volume. Taking $\theta_D = 500$ K,[14] results in a $l_{ph}$ that decreases from 30.4 Å for $x = 0$ to 6.8 Å for $x = 0.8$. This reduction agrees with the emergence of a glassy state in which $l_{ph}$ is expected to be of the order of the inter-atomic spacing (c.f. $l_{ph} \approx 7$ Å for a variety of $SiO_2$ glasses calculated using the same model).[15]

In order to assess the impact of A-site vacancies on the power factor a Nb substituted $Sr_{1-x}La_{0.67x}\square_{0.33x}Ti_{0.8}Nb_{0.2}O_{3-\delta}$ series was prepared. The temperature dependence of S and the electrical resistivity ($\rho = 1/\sigma$) are given in Fig. S3. The S(T) remain linear but are reduced to $S_{300K}$ = -25 μV K$^{-1}$ and $S_{825K}$ = -80 μV K$^{-1}$ for x = 0.4 (13% □) and 0.8 (27% □). The resistivity for x = 0 has a metallic temperature dependence with $\rho_{300K}$ = 1.1 mΩ cm and $\rho_{825K}$ = 3.6 mΩ cm. The x = 0.4 and 0.8 samples have similar resistivities (e.g. $\rho_{825K}$ ~ 4 mΩcm) and show a transition from semiconducting to metallic behaviour at 650-700 K. The presence of the semiconducting to metallic transition suggests that the A-site vacancies result in an additional activation barrier which does not affect the high-temperature transport. The introduction of vacancies therefore results in a decrease in S, the presence of a semiconductor-metal transition, while similar resistivities are observed. This results in reduced power factors $S^2/\rho$ = 0.2 mW m$^{-1}$ K$^{-2}$ at 825 K for x = 0.4 and x = 0.8 (compared to 0.6 mW m$^{-1}$ K$^{-2}$ for x = 0).

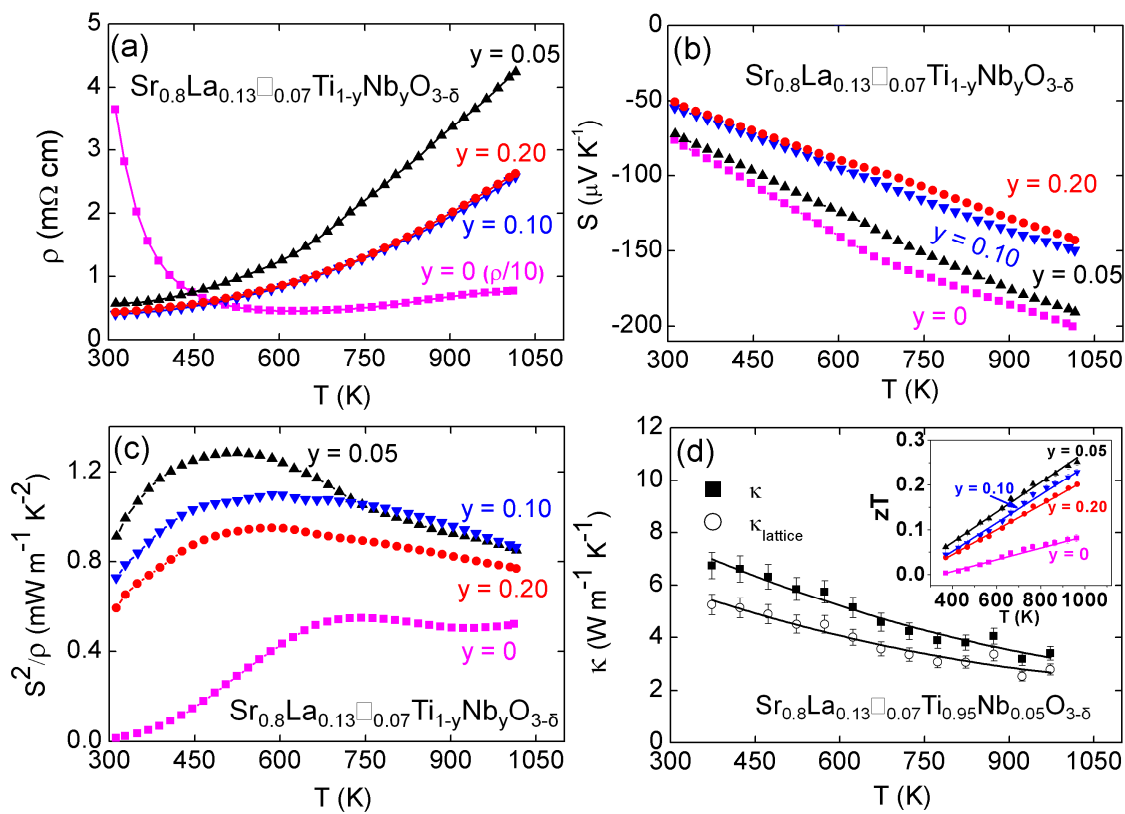

**Fig. 2.** Thermoelectric performance of Nb substituted SrTiO$_3$ with 7% A-site vacancies. Temperature dependence of (a) the electrical resistivity (ρ), (b) the Seebeck coefficient (S), (c) the thermoelectric power factor ($S^2/\rho$) for the Sr$_{0.80}$La$_{0.13}$□$_{0.07}$Ti$_{1-y}$Nb$_y$O$_{3-\delta}$ series, and (d) the total (κ) and lattice ($\kappa_{lat}$) thermal conductivity for x = 0.2. The inset shows the estimated temperature dependence of the figure of merit, ZT, for all samples. The ρ of the y = 0 sample has been divided by 10.

Interestingly samples with x = 0.2 (7% □) were consistently found to have a low ρ, leading to large power factors $S^2/\rho$ = 1 mW m$^{-1}$ K$^{-2}$. For this reason, these compositions were optimised by variation of the Nb content. This yielded the Sr$_{0.80}$La$_{0.13}$□$_{0.07}$Ti$_{1-y}$Nb$_y$O$_{3-\delta}$ series (0 ≤ y ≤ 0.2) whose properties are summarised in Fig. 2. The ρ(T) curves for y ≥ 0.05 have a metallic temperature dependence with $\rho_{300K}$ ≈ 0.5 mΩcm and $\rho_{1000K}$ = 2.5-4 mΩcm, while the y = 0 sample is semiconducting. The S(T) are linear and range from -50 ≤ $S_{300K}$ ≤ -75 µV K$^{-1}$ to -140 ≤ $S_{1000K}$ ≤ -200 µV K$^{-1}$. This results in large $S^2/\rho$ values up to ~1.3 mW m$^{-1}$ K$^{-2}$ for y = 0.05. Measurement of κ(T) reveals a conventional 1/T dependence, in keeping with the results presented in Fig. 1. The lattice contribution dominates κ(T) and decreases from 5.5 W m$^{-1}$ K$^{-1}$ (323 K) to 2.5 W m$^{-1}$ K$^{-1}$ (1000 K). Calculation of the figure of merit leads to a maximum ZT ≈ 0.3 at 1000 K for y = 0.05, which is comparable to the best reported values for non A-site deficient SrTi$_{1-y}$Nb$_y$O$_3$ and La$_{1-x}$Sr$_x$TiO$_3$ based compositions.[4, 5]

Neutron powder diffraction was used to confirm the A-site deficiency, and to investigate the oxygen stoichiometry of the best performing sample (Table S2, Table S3, and Fig. S4). The refined composition was found to be Sr$_{0.798(3)}$La$_{0.130(3)}$Ti$_{0.95}$Nb$_{0.05}$O$_{2.91(3)}$, which confirms that the A-site vacancy is maintained under reducing conditions, and highlights the presence of ~3% oxygen vacancies. The sub-stoichiometry in the oxygen sublattice reduces the average oxidation state of the transition metal cations (from +4 to +3.83), *i.e.* increases the electron concentration (from nominally 0.05 to 0.23 e$^-$ per transition metal). This interplay between oxygen stoichiometry and Nb content is also evident from Fig. 2a-c that reveals only modest changes in electronic transport for y ≥ 0.05, suggesting that y and δ compensate to maintain similar carrier concentrations.

The most striking feature of the data presented is that κ(T) changes from being characteristic of a well ordered solid to that of a glass. This can be achieved in a highly controllable fashion by introducing vacancies on the perovskite A-site. Figure 3 shows a phase diagram that illustrates the vacancy dependence of the thermal conductivity for the samples presented in this manuscript. This shows that the conducting x = 0.2 sample (Fig. 2d) fits in well with the trends established by the oxygen stoichiometric samples presented in Fig. 1. From the literature it is known that the vacancies are initially randomly distributed (x ≤ 0.3) and subsequently long-range order in vacancy rich layers within the perovskite structure (x ≥ 0.83).[11] The observed κ = 2.5 W m$^{-1}$ K$^{-1}$ approaches Cahill et al.'s minimum thermal conductivity for SrTiO$_3$, which was estimated to be 1.4 W m$^{-1}$ K$^{-1}$ at 300 K and 1.8 W m$^{-1}$ K$^{-1}$ at 1000 K.[16] The introduction of mixtures of Sr, La and vacancies will result in significant mass fluctuations and microstrain, which are effective disruptors of lattice vibrations. For alloying on a single crystallographic site the phonon relaxation time is given by:[17]

$$\tau_{PD}^{-1} = \frac{V\omega^4}{4\pi v_p^2 v_g} \left( \sum_i f_i \left(1 - \frac{m_i}{m_{av}}\right)^2 + \sum_i f_i \left(1 - \frac{r_i}{r_{av}}\right)^2 \right)$$

Where V is the volume per atom, ω is the phonon frequency, $v_p$ ($v_g$) is the phonon phase (group) velocity, $f_i$ is the fraction of atoms with mass $m_i$ and radius $r_i$ that occupy the site with average mass ($m_{av}$) and radius ($r_{av}$). This shows that both mass and size differences contribute to reducing the phonon mean free path ($l_{ph} \propto \tau_{PD}$). The size of an A-site vacancy can be estimated from the □-O bond distance, and is comparable to the ionic radii of Sr$^{2+}$ and La$^{3+}$, leading to a modest increase in the size mismatch term. In contrast, the zero mass of the vacancy will lead to a large mass fluctuation term. For Sr$_{0.2}$La$_{0.53}$□$_{0.27}$TiO$_3$, the mass difference sum equals 0.4, while a 50/50 mix of Sr and La (Sr$_{0.5}$La$_{0.5}$TiO$_3$) yields only 0.05.

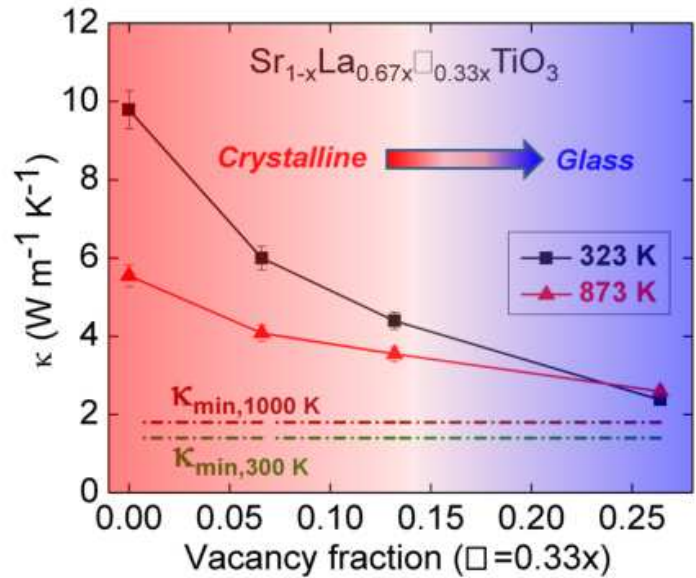

**Fig. 3**. Vacancy concentration dependence of the lattice thermal conductivity at 323 and 873 K for the investigated A-site deficient $SrTiO_3$ samples. The calculated minimum thermal conductivity for $SrTiO_3$ using Cahill et al's model is also shown.[16] The colour gradation indicates the gradual transition from crystalline to glass-like behaviour.

This suggests a large reduction of $l_{ph}$ may be expected compared to compositions not containing vacancies. However, it is not apparent that a vacancy will scatter phonons in the same manner as an atom.[16] If the vacancies are not included in the calculation, the mass difference sum (0.02) is comparable to the A- and B-site doped compositions in the literature, and therefore not able to explain the observed glassy $\kappa(T)$.

The glass-like thermal conductivity in relaxor ferroelectrics and $Nd_{0.67-x}Li_{3x}TiO_3$ has been linked to phase segregation,[13] but there is no evidence for this for our compositions (Fig. S1 and Ref. 11). Another explanation is that the vacancies result in a low energy "rattling" vibration mode that dissipates heat, leading to a glassy thermal conductivity in a crystalline material.[17] For the A-site deficient perovskites the formation of A-O-□ linkages could lead to soft A-O bonds, and the emergence of a rattling mode and phonon-glass state. The introduction of vacancies in $SrTiO_3$ results in a reduction of S. This decreases $S^2/\rho$, yielding an estimated ZT = 0.07 at 825 K for samples with 27% vacancies. If power factors of 1-2 mW m$^{-1}$ K$^{-2}$ can be maintained ZT = 0.3-0.6 is achievable. This would be a significant improvement over the current state-of-the-art. The similar ρ values compared to samples without vacancies suggest that there is scope for further optimisation. The power factors $\leq 1.3$ mW m$^{-1}$ K$^{-2}$ for samples with 7% vacancies are high for polycrystalline samples, and they are easier to carrier dope than the analogous SrTi$_{1-y}$Nb$_y$O$_{3-\delta}$ compositions. This is beneficial for large-scale preparation and for long-term stability under operating conditions.

**Conclusions:**

To conclude, we have demonstrated that A-site vacancies can be used to induce a glass-like thermal conductivity. This provides a promising route for further optimisation of the thermoelectric performance of SrTiO$_3$ based thermoelectric materials.

**References:**

1. G. J. Snyder and E. S. Toberer, *Nat. Mat.*, 2008, **7**, 105; J. R. Sootsman, D. Y. Chung and M. G. Kanatzidis, *Angew. Chem., Int. Ed.*, 2009, **48**, 8616.

2. J. W. Fergus, *J. Eur. Ceram. Soc.,* 2012, **32**, 525; J. He, Y. F. Liu and R. Funahashi, *J. Mater. Res.,* 2011, **26**, 1762; K. Koumoto, I. Terasaki and R. Funahashi, *MRS Bulletin*, 2006, **31**, 206.

3. S. Funahashi, T. Nakamura, K. Kageyama and H. Leki, *J. Appl. Phys.*, 2011, **109**, 124509.

4. S. Ohta, T. Nomura, H. Ohta, M. Hirano, H. Hosono and K. Koumoto, *Appl. Phys. Lett.*, 2005, **87**, 092108; T. Okuda, K. Nakanishi, S. Miyasaka and Y. Tokura, *Phys. Rev. B,* 2001, **63**, 113104; M. Yamamoto, H. Ohta and K. Koumoto, *Appl. Phys. Lett.*, 2007, **90**, 072101.

5. S. Ohta, T. Nomura, H. Ohta and K. Koumoto, *J. Appl. Phys.*, 2005, **97**, 034106.

6. H. Muta, K. Kurosaki and S. Yamanaka, *J. Alloys Compd.,* 2005, **392**, 306.

7. G. D. Mahan, in *Solid State Physics, Vol 51: Advances in Research and Applications*, Vol. 51 (Eds: H. Ehrenreich, F. Spaepen), 1998, 81.

8. K. Kato, M. Yamamoto, S. Ohta, H. Muta, K. Kurosaki, S. Yamanaka, H. Iwasaki, H. Ohta and K. Koumoto, *J. Appl. Phys.*, 2007, **102**, 116107.

9. A. V. Kovalevsky, A. A. Yaremchenko, S. Populoh, A. Weidenkaff and J. R. Frade, *J. Phys. Chem.* C. 2014, **118**, 4596.

10. H. Kawakami, M. Saito, H. Takemoto, H. Yamamura, Y. Isoda and Y. Shinohara, *Mat. Trans.*, 2013, **54**, 1818.

11. P. D. Battle, J. E. Bennett, J. Sloan, R. J. D. Tilley and J. F. Vente, *J. Solid State Chem.,* 2000, **149**, 360; C. J. Howard, G. R. Lumpkin, R. I. Smith and Z. M. Zhang, *J. Solid State Chem.,* 2004, **177**, 2726.

12 T. M. Tritt, Ed. *Thermal Conductivity: Theory, Properties and Applications*, Kluwer Academic, New York 2004.

13. M. Tachibana, T. Koladiazhnyi and E. Takayama-Muromachi, *Appl. Phys. Lett.*, 2008, **93**, 092902; Y. Ba, C. Want, Y. Wang, N. Norimatsu, M. Kusunoki and K. Koumoto. *Mat. Lett.* 2013, **97**, 191.

14. M. Ahrens, R. Merkle, B. Rahmati and J. Maier, *Phys. B-Cond. Matt.,* 2007, **393**, 239.

15. C. Kittel, *Phys. Rev.*, 1949, **75**, 972.

16. D. G. Cahill, S. K. Watson and R. O. Pohl, *Phys. Rev. B*, 1992, **46**, 6131. Y. F. Wang, K. Fujinami, R. Z. Zhang, C. L. Wan, N. Wang, Y. S. Ba and K. Koumoto, *Appl. Phys. Exp*, 2010, **3**, 031101.

17. E. S. Toberer, A. Zevalkink and G. J. Snyder, *J. Mater. Chem.,* 2011, **21**, 15843.

18. A. C. Larson and R. B. Von Dreele, *General Structure Analysis System (GSAS)* 2000; B. H. Toby, *Journal of Applied Crystallography*, 2001, **34**, 210.

Supplementary information for:

# Glass-like thermal conductivity in SrTiO$_3$ thermoelectrics induced by A-site vacancies

S. R. Popuri,[a] A. J. M. Scott,[a] R. A. Downie,[a] M. A. Hall,[a] E. Suard,[b] R. Decourt,[c] M. Pollet,[c] and J.-W. G. Bos[a]*

[a] Institute of Chemical Sciences and Centre for Advanced Energy Storage and Recovery, School of engineering and physical sciences, Heriot-Watt University, Edinburgh, EH14 4AS, United Kingdom. *j.w.g.bos@hw.ac.uk

[b] Institut Laue-Langevin, Grenoble, F-38000, France

[c] CNRS, Université de Bordeaux, ICMCB, 87 avenue du Dr. A. Schweitzer, Pessac F-33608, France

**Table S1.** Lattice parameters from Rietveld fits against laboratory X-ray data and pellet densities for the Sr$_{1-x}$La$_{0.67x}$□$_{0.33x}$TiO$_3$, Sr$_{1-x}$La$_{0.67x}$□$_{0.33x}$Ti$_{0.80}$Nb$_{0.20}$O$_{3-\delta}$ and Sr$_{0.8}$La$_{0.13}$□$_{0.07}$Ti$_{1-y}$Nb$_y$O$_{3-\delta}$ series.

| x | y | a (Å) | Density (%) |
|---|---|---|---|
| 0 | 0 | 3.9052(1) | 95(1) |
| 0.4 | 0 | 3.8966(1) | 97(1) |
| 0.8 | 0 | 3.8843(1) | 92(1) |
| 0 | 0.2 | 3.9267(1) | 92(1) |
| 0.4 | 0.2 | 3.9056(1) | 90(1) |
| 0.8 | 0.2 | 3.8935(1) | 91(1) |
| 0.2 | 0 | 3.9082(1) | 90(1) |
| 0.2 | 0.05 | 3.9135(1) | 91(1) |
| 0.2 | 0.10 | 3.9163(1) | 95(1) |
| 0.2 | 0.2 | 3.9341(1) | 93(1) |

Space group Pm-3m, Sr/La/□ (0, 0, 0), Ti/Nb (1/2, 1/2, 1/2), O (1/2, 1/2, 0).

**Table S2.** Structural parameters for $Sr_{0.8}La_{0.13}Ti_{0.95}Nb_{0.05}O_{2.91(3)}$ from a Rietveld fit against super-D2B neutron powder diffraction data.

|       | Wyckoff | x         | y         | z    | Occupancy           | $U_{iso}$ (Å$^2$) |
|-------|---------|-----------|-----------|------|---------------------|-------------------|
| Sr/ La | 4b     | 0         | 0.5       | 0.25 | 0.798(3)/ 0.130(3)  | 0.006(1)          |
| Ti/ Nb | 4c     | 0         | 0         | 0    | 0.95/ 0.05          | 0.005(1)          |
| O1    | 4a      | 0         | 0         | 0.25 | 0.96(5)             | 0.012(1)          |
| O2    | 8h      | 0.2381(2) | 0.7381(2) | 0.5  | 0.97(2)             | 0.006(1)          |

Space Group I4/mcm; a = 5.5327(1) Å; c = 7.8324(4) Å.

**Table S3.** Selected bond distances (Å) and angles (°) for $Sr_{0.8}La_{0.13}Ti_{0.95}Nb_{0.05}O_{2.91(3)}$.

|                      | Distance/ Angle |
|----------------------|-----------------|
| Ti/Nb-O1 (×2)        | 1.95809(9)      |
| Ti/Nb-O2 (×4)        | 1.95831(9)      |
|                      |                 |
| Ti/Nb-O1-Ti/Nb       | 180             |
| Ti/Nb-O2-Ti/Nb       | 174.57(9)       |
|                      |                 |
| Sr/La-O1 (×4)        | 2.76635(7)      |
| Sr/La-O2 (×4)        | 2.7030(10)      |
| Sr/La-O2 (×4)        | 2.8341(11)      |

**Fig. S1.** Room temperature powder X-ray diffraction patterns for the (a) $Sr_{1-x}La_{0.67x}\square_{0.33x}O_3$, (b) $Sr_{1-x}La_{0.67x}\square_{0.33x}Ti_{0.8}Nb_{0.2}O_{3-\delta}$ and (c) $Sr_{0.80}La_{0.13}\square_{0.07}Ti_{1-y}Nb_yO_{3-\delta}$ series.

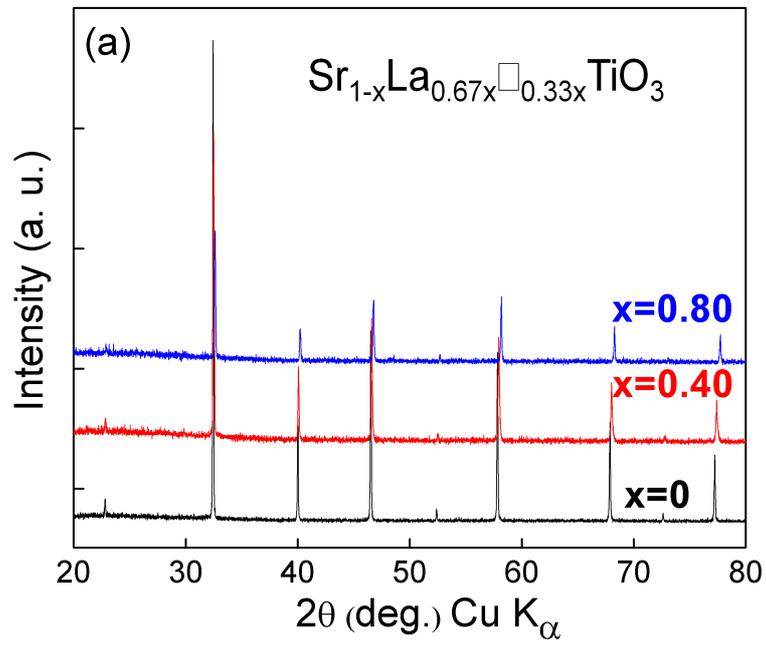

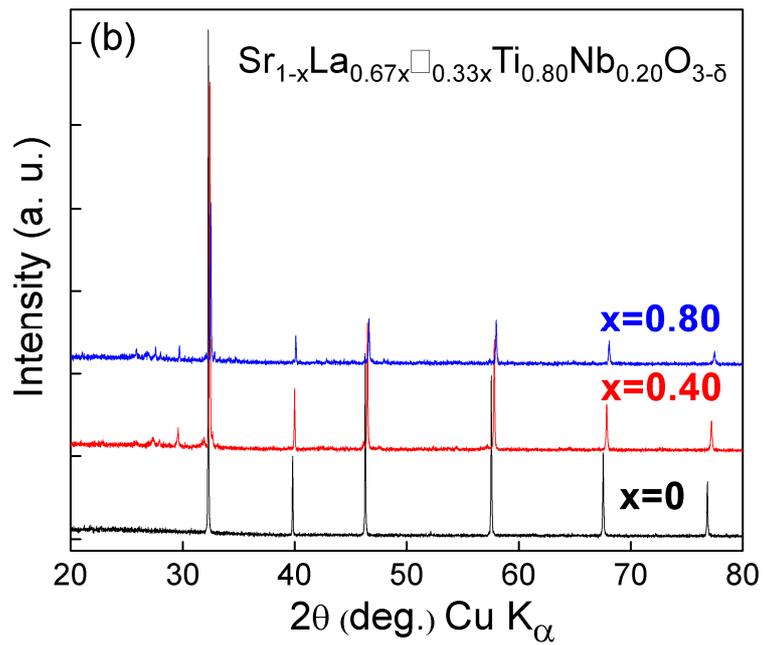

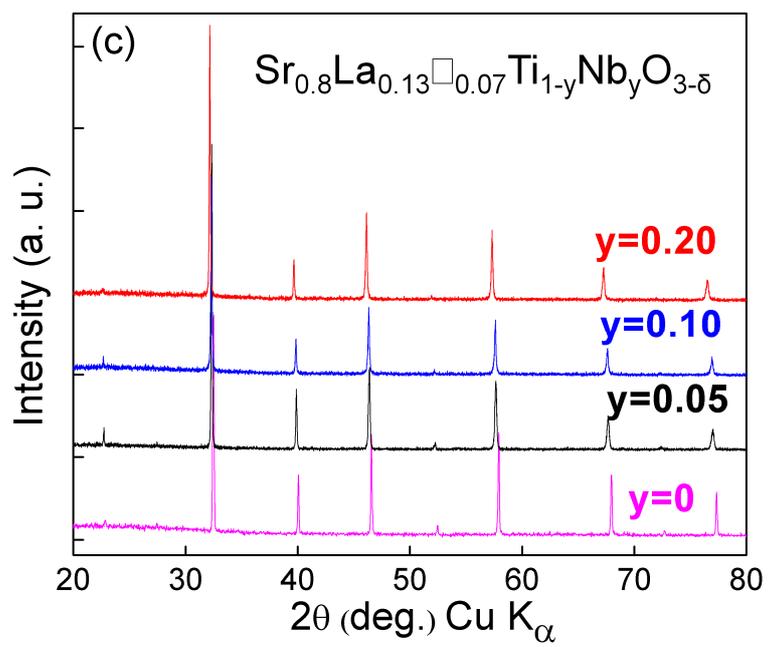

**Fig. S2.** Temperature dependence of the thermal diffusivity (α) and specific heat ($C_p$) for the $Sr_{1-x}La_{0.67x}\square_{0.33x}TiO_3$ series.

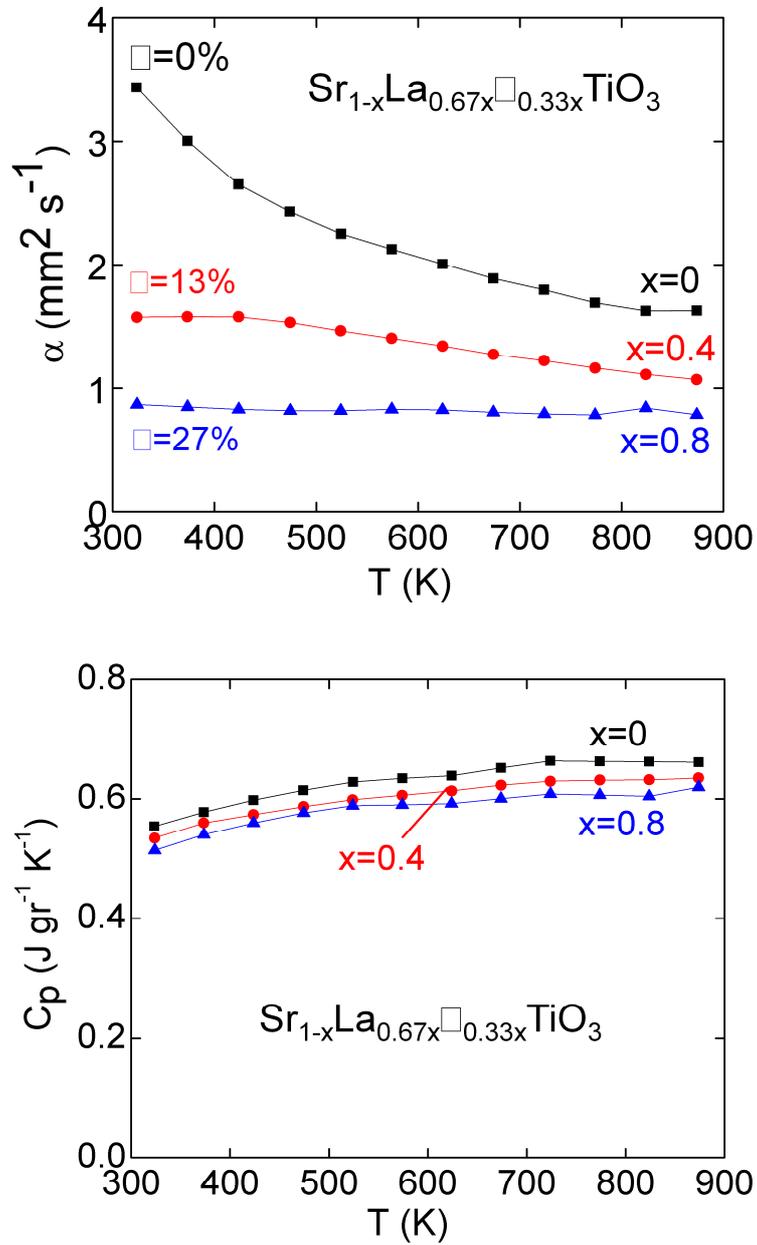

**Fig. S3**. Temperature dependence of the Seebeck coeffient (S) and electrical resistivity (ρ) for the Sr$_{1-x}$La$_{0.67x}$□$_{0.33}$Ti$_{0.8}$Nb$_{0.2}$O$_3$ series.

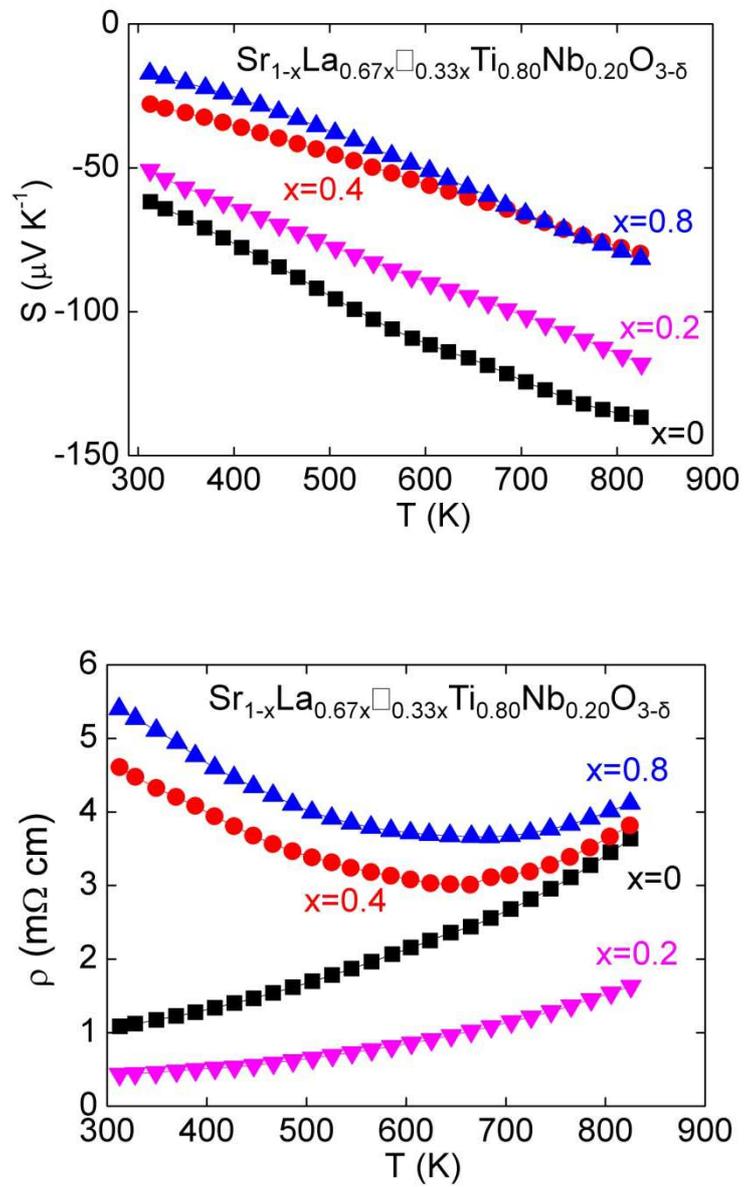

**Fig. S4.** Observed (circles), calculated (solid line) and difference Rietveld neutron diffraction profiles for $Sr_{0.80}La_{0.13}\square_{0.07}Ti_{0.95}Nb_{0.05}O_{2.91(3)}$. The bottom row of Bragg markers is for a 1.0(2) wt% $TiO_2$ impurity. Fit statistics: $\chi^2 = 4.1$, $wR_p = 4.0\%$, $R_p = 3.2\%$, $R_F^2 = 2.2\%$.

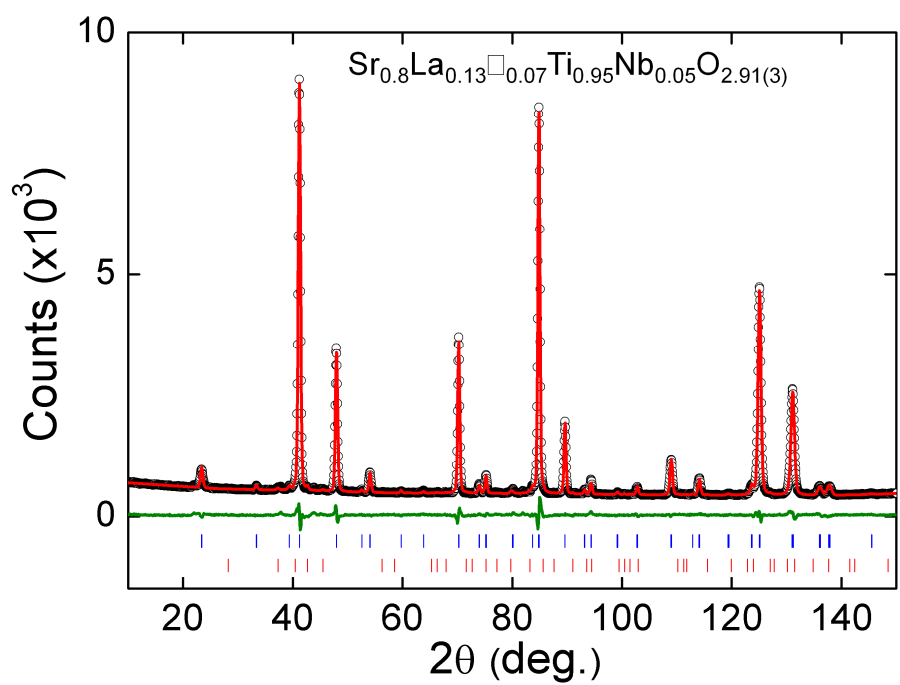